\documentclass{eas}
\usepackage{graphicx}
\def\spose#1{\hbox to 0pt{#1\hss}}
\def\lta{\mathrel{\spose{\lower 3pt\hbox{$\mathchar"218$}}
     \raise 2.0pt\hbox{$\mathchar"13C$}}}
\def\gta{\mathrel{\spose{\lower 3pt\hbox{$\mathchar"218$}}
     \raise 2.0pt\hbox{$\mathchar"13E$}}}

\def\=#1{\overline{#1}}

\def\deg{^\circ}             

\def\kms{{\rm\,km\,s^{-1}}}

\def\kpc{{\rm\,kpc}}

\def\msun{{\rm\,M_\odot}}

\def\gyr{{\rm\,Gyr}}

\begin{document}

\TitreGlobal{Mass Profiles and Shapes of Cosmological Structures}

\title{Mass Distribution and Bulge Formation in the Milky Way Galaxy}
\author{Ortwin Gerhard}\address{Astronomisches Institut, Universit\"at
Basel, Venusstrasse 7, CH-4102 Binningen, Switzerland; current address: 
Max-Planck-Institut f\"ur Extraterrestrische Physik, Giessenbachstrasse,
D-85748 Garching, Germany}
%
\runningtitle{Mass Distribution and Bulge Formation in the Galaxy}
\setcounter{page}{1}
\index{Gerhard, O.E.}


%
\begin{abstract}
In its first part, this paper summarizes recent work on the mass and
shape of the Galactic dark halo. The second part presents a review of
the large-scale structure of the Milky Way, and of the evidence that
the inner Galaxy is dominated by baryonic matter. This is briefly
compared with the predictions of $\Lambda$CDM and MOND. Finally, a
summary is given of bulge formation from clumpy, gas-rich disks, a
process which may give rise to old, disk-like, $\alpha$-rich bulges
similar to the Galactic bulge.
\end{abstract}
\maketitle
%
\section{Introduction}

Our position inside the Milky Way has both advantages and
drawbacks. On the one hand, a number of experiments are possible that
give complementary information on Galactic structure to
that attainable for external galaxies. Notable examples are
microlensing surveys, three-dimensional velocity data, local dynamical
and stellar population measurements, determination of line-of-sight
density distributions, etc. On the other hand, our view from inside
the disk is often obscured, distance measurements are often needed but
difficult, and deciphering the large-scale structure of the Galaxy is
a much more involved process than from the image of an external
galaxy. Both aspects are evident in the following sections: in \S2, 
we review recent work on measuring the radial mass distribution and
shape of the Galactic dark matter halo. Then in \S3, we summarize
constraints on the distribution of baryonic matter in the Milky Way,
based on NIR, gas kinematic, and microlensing data. Finally in
\S4, we discuss a model of bulge formation from clumpy gas-rich disks
which has some promise for understanding Galactic bulge observations.

\section{The Galactic Dark Halo}

\subsection{Total Halo Mass and Radial Mass Distribution}

HI radial velocity measurements constrain the Galactic rotation curve
to radii $\lta 20\kpc$ (Honma \& Sofue 1997). Determinations of the
total mass of the Galaxy are therefore based on the kinematics of
globular clusters, satellite galaxies, and distant halo stars.  There
are not many of these; so further assumptions are needed.  Based on
radial velocities of 27 globular clusters and satellites at
galactocentric $R>20\kpc$, and additional proper motions for six of
these, Wilkinson \& Evans (1999) estimated the Milky Way's mass inside
$50\kpc$ to be $M(50\kpc)=5.4^{+0.2}_{-3.6}\times 10^{11} \msun$, and
its total mass as $M_t=1.9^{+3.6}_{-1.7}\times 10^{12} \msun$. They
used a Baysian likelhood method, following earlier work by Little \&
Tremaine (1987) and Kochanek (1996), in which they assumed a spherical
halo mass model with a truncation radius, and a particular ansatz for
an anisotropic distribution function, assumed to describe the entire
ensemble of tracers. Then the probability of the tracers to have their
observed positions and velocities in this model is determined, and
maximized over the model's total mass and constant anisotropy.

Sakamoto et al.\ (2003) applied this method to a larger sample of
tracer objects, including 413 FHB halo stars, half of them with proper
motions. Their estimate for $M_t=2.5^{+0.5}_{-1.0}\times 10^{12}
\msun$ is somewhat larger, while they obtain
$M(50\kpc)=5.5^{+0.0}_{-0.2}\times 10^{11} \msun$, very similar to the
earlier value but with much smaller errors. In my understanding this
illustrates the limitations of this approach: the parametric model is
seemingly very tightly constrained by the large number of new data
points at mostly small radii; it is not flexible enough to adjust to
these data locally while keeping the mass at large radii constrained
by mostly the outer data points only. This suggests that these error
bars should be viewed with caution.

Recently, Battaglia et al.\ (2005) have compiled a new halo tracer
sample including distant halo giant stars with accurate distances and
radial velocities. From this they determine more accurately the radial
velocity dispersion profile (constant at $\sim120\kms$ out to
$30\kpc$, then declining to $\sim50\kms$ at $120\kpc$). They then
compare with dispersion profiles predicted from the Jeans equation for
specified halo mass distribution and anisotropy. The data rule out an
isothermal halo with constant anisotropy, but can be explained either
by a dark matter halo truncated at large radii and constant velocity
anisotropy, or by a NFW halo with less steep density profile whose
velocity anisotropy becomes tangential at large radii. The total halo
mass determined in these cases is $M_t=1.2^{+1.8}_{-0.5}\times 10^{12}
\msun$ and $M_v=0.8^{+1.2}_{-0.2}\times 10^{12} \msun$, respectively.
These values are smaller by a factor $\sim 2$ than the earlier total
mass estimates, but are still consistent within the joint error
ranges. For comparison, the favoured $\Lambda$CDM model from Klypin et
al.\ (2002) has $M_v=1.0\times 10^{12} \msun$.

\subsection{Halo Shape}

Until recently, data constraining the shape of the Galactic halo were
hard to come by. Olling \& Merrifield (2000) found evidence for a
near-spherical oblate halo with shortest-to-longest axis ratio
$q_\rho=0.8$, based on two techniques: (1) by comparing the mean halo
density inferred from the rotation curve with the measured mass column
density near the Sun, and (2) by comparing the predicted flaring of
the gas disk in various flattened halo models with the measured
flaring of the HI layer. However, these methods only agreed for
somewhat small values of the Galactic constants $R_0$ and $V_0$; for
more standard values, the second method would actually favour a
prolate halo, in disagreement with the first.

The detailed mapping of the Sagittarius stream and the associated
radial velocity measurements are bringing a significant advance to
this subject. From the observation that the Carbon stars in the Sgr
stream are consistent with a great circle on the sky, Ibata et al.\
(2001) argued that the precession rate must be slow and hence the halo
nearly spherical.  The recent all-sky map of the Sgr stream as traced
by the M stars in the 2MASS survey (Majewski et al.\ 2003), together
with other detections of the tidal streams of the Sgr galaxy
(Mart\'inez-Delgado et al.\ 2004), have provided strong constraints on
the orbit of the tidal debris in the halo potential, and hence on the
halo shape. From their simulations, Mart\'inez-Delgado et al.\ (2004)
obtain a halo potential flattening $q_\Phi=0.85$. More recent fitting
of the orbital planes (poles) of the Sgr leading and trailing debris
in the 2MASS M giant data gives a very similar result (Johnston et
al.\ 2005). However, Helmi (2004) finds that simulations of Sgr debris
best agree with radial velocity measurements for part of the leading
and trailing streams if the halo potential is prolate, with
$q_\Phi=1.25$. Radial velocities of course not only tell us about the
speeding up of the stars as they fall towards the Galactic plane,
which is different in oblate and prolate potentials. They also depend
on how the orbit changes direction at the same time.  It remains to be
seen whether the positional and kinematic analyses of the Sgr stream
can be reconciled for a triaxial halo potential, such as is indicated
by SDSS starcounts for the {\sl stellar} halo (Newberg \& Yanny 2005).

\section{Structure and Radial Mass Distribution of the Galaxy}

\subsection{Large-Scale Structure}

The Milky Way is a barred galaxy with fairly weak outer spiral
structure. Modern models of the luminosity distribution are based on
NIR data, both from integrated light observations (COBE/DIRBE) and
starcounts in the NIR. Bissantz \& Gerhard (2002, hereafter BG)
presented non-parametric models of the Galactic luminosity density
interior to the Sun. Like several previous models (e.g., Binney,
Gerhard \& Spergel 1997) they are based on the COBE/DIRBE photometry,
but also include a spiral arm model after Ortiz \& L\'epine (1993) and
were verified a posteriori with the line-of-sight distributions of
clump giant stars in several bulge fields (Stanek et al.\ 1997). The
best models were obtained for bar angles $20-25\deg$ relative to the
Sun-Galactic Centre line and had a thin bar (10:$\sim\!3.5$:3) with
length $\sim\!3.5\kpc$ and a short disk scale-length, $\sim\!2.1
\kpc$. The latter is somewhat shorter than the disk scale-length
obtained from DENIS ($2.5\kpc$, Robin et al.\ 2003) and 2MASS
($\sim\!3\kpc$, L\'opez-Corredoira et al.\ 2004) NIR starcounts.  The
2MASS analysis also indicates that the disk profile flattens out
inside $\sim\!4\kpc$, rather than rising exponentially into the
center. After subtracting the disk starcounts, inversion of the 2MASS
counts results in a boxy bulge with slightly larger axis ratios than
in the BG model (10:$\sim\!5$:4; L\'opez-Corredoira, Cabrera-Lavers \&
Gerhard 2005).  Recently, Babusiaux \& Gilmore (2005) obtained
distance distributions of clump giant stars from deep NIR data in
several bulge fields, which are consistent with a triaxial bar with
in-plane axis ratio 10:3-4, bar angle $22\pm 5.5\deg$, extending to
$\sim\!2.5\kpc$, and circumscribed by an inner ring.  Thus a fairly
consistent picture of the inner Galaxy has emerged, with perhaps the
main uncertainy being the density distribution of the disk inside a
few kpc.

Taking the BG model with constant NIR mass-to-light ratio as a mass
model, Bissantz, Englmaier \& Gerhard (2003) simulated the Galactic
gas flow in the Galactic gravitational potential. Comparing simulated
$(l,v)$-plots with the observed CO $(l,v)$-diagram they determined the
pattern speeds of the Galactic bar and spiral arms. The observed
$(l,v)$-plot is best reproduced if the bar corotation radius is
$\sim\!3.5\kpc$, consistent with Dehnen's (2000) independent
determination of the pattern speed, while the spiral arm pattern speed
is substantially lower, with corotation radius near the Sun. The
spiral arms are relatively weak in luminosity and one pair may we
weaker than the other (Drimmel \& Spergel 2001), but the
$(l,v)$-diagram is best reproduced if all four spiral arms have mass.

\subsection{Mass of the Bulge and Disk}

Scaling the simulated terminal velocity curve of this gas flow model,
a rather good fit to the observed HI and CO terminal velocity curve
can be obtained for $l\lta 50\deg$ (Englmaier \& Gerhard 1999,
Bissantz et al.\ 2003).  The model with this scaling, which is
equivalent to a maximum disk model in external galaxies, predicts a
circular velocity of $v_c=190-200\kms$ at galactocentric radii
$2-4\kpc$, and a circular velocity at the solar radius $v_c(R_0)\simeq
183\kms$, $\sim 40\kms$ lower than the assumed $v_c(R_0)=220\kms$.
With a similar scaling, a self-consistent N-body model matched to the
BG luminosity model (Bissantz, Debattista \& Gerhard 2004) also
reproduces the radial velocity and proper motion velocity dispersions
in several bulge fields. With a reasonable mass function for the bulge
and disk stars, this dynamical model in addition accounts for the
distribution of microlensing event durations determined by the MACHO
collaboration.

In the Milky Way, the degeneracy in such analysis between dark and
luminous matter can be lifted -- the mass-to-light ratio obtained for
the disk and bulge stars from the dynamical fits can be independently
verified with microlensing observations. BG give an optical depth map
for clump giant stars in their model with maximum M/L. For the clump
giant stars the most secure optical depths can be measured; the
published values to date are, in units of $10^{-6}$,
$\tau_{-6}=2.17^{+0.47}_{-0.38}$ at $(l,b)=(1.5\deg, -2.68\deg)$
(MACHO collaboration, Popowski et al.\ 2005), and
$\tau_{-6}=0.96\pm0.3$ at $(l,b)=(2.5\deg, -4\deg)$ (EROS
collaboration, Afonso et al.\ 2003). The BG model predicts
$\tau_{-6}=2.4$ and $\tau_{-6}=1.2$ at these positions, slightly
higher but within the errors. More recent EROS data on $\tau_{-6}$ at
three points in latitude (Hamadache et al.\, in preparation) also is
in good agreement with the BG model.  Thus the observed clump giant
optical depths now agree quantitatively with the model predictions,
for a model in which the inner Galaxy is dominated by baryonic
mass. There is no room for dynamically significant dark matter in the
inner Galaxy (see also Binney, Bissantz \& Gerhard 2000).

\subsection{Comparison with $\Lambda$CDM and MOND}

Milky Way mass models based on disk formation with adiabatic
contraction in cuspy dark matter halos in the $\Lambda$CDM cosmology
were investigated by Klypin et al.\ (2002). Their favoured model A1
predicts a circular velocity at $R=2-4\kpc$ of $\sim\!150\kms$,
significantly lower than the $190-200\kms$ predicted by models for the
combined NIR, gas dynamical and microlensing data. More consistent
with these data is their model B1, which predicts $v_c\!\sim\!185\kms$
at $R=2-4\kpc$. Model B1 differs from A1 in that a mechanism is
assumed by which the disk loses a factor $1.5-2$ of its angular
momentum to the halo during the collapse. It is also more akin to a
maximum disk model and has a larger disk mass than model A1.

Famaey \& Binney (2005) have investigated the predictions of
the BG model in MOND dynamics. They find that with a suitable but
non-standard choice of the MOND interpolating function, the standard
acceleration constant $a_0$, and a realistic value for the disk M/L,
the Galactic terminal curve can be fitted. With the standard
interpolating function some dark matter would still be required if
$v_c(R_0)=220\kms$, but not for $v_c(R_0)=200\kms$.  However, it may
be premature to conclude that MOND can explain the Milky Way dynamics,
because of the evidence cited above for a triaxial distribution of
spheroid stars (Newberg \& Yanny 2005). The SDSS starcounts indicate
that the major axis of the spheroid is approximately perpendicular to
that of the Galactic bar. The spheroid itself, however, contributes
little to the Galactic rotation curve near and outside the solar
radius, an elongated disk does not support its elongated orbits, and
so the most likely source of triaxial potential would have to be a
massive dark halo after all.

\section{Origin of the Milky Way Bulge}

\subsection{Old, $\alpha$-rich Galactic Bulge Stars and
Common Routes to Bulge Formation}

The two scenarios commonly discussed for bulge formation are (1) the
merging of early disks and fragments, and (2) the secular evolution of
disks and bars. The former typically creates large bulges unlike that
of the Milky Way, and the latter is likely to create solar abundance,
intermediate-age bulges. This is because an extended history of star
formation has already enriched the gas before the disk surface density
grows large enough to trigger the bar and buckling instabilities.

The Galactic bar does not obviously fit into these scenarios. On the
one hand, it is a rapidly rotating bar, so likely descends from the
disk. On the other hand, the stars in Baade's window are old, $\sim
12\gyr$ (e.g., Ortolani et al.\ 1995), and $\alpha$-rich (Mc William
\& Rich 1994, Rich \& Oviglia 2005). One possibility is that the
inner disk formed rapidly and the bulge bar formed from the
disk, all before the iron from SN type IA could significantly enrich the
gas. Another is the early formation of an $\alpha$-rich bulge,
followed by the later addition of a barred bulge component from the
then $\sim$solar abundance disk. In this case one expects abundance
ratio gradients.  In the following, a mechanism is described by which
an $\alpha$-rich bulge component forms from a clump-unstable,
star-bursting disk.  High-redshift morphological observations suggest
that this mechanism may indeed be at work in young late-type disk
galaxies, and may thus be of relevance for the origin of the Galactic
bulge.

\subsection{Clump Instability in Gas-Rich Disks}

It is well-known that cold gas has a destabilizing influence on
galactic disks. Immeli et al.\ (2004a) studied this in the context of
a chemo-dynamical model including a dark halo, stars, and a two-phase
interstellar medium with feedback processes from the stars. When most
baryons are still in the form of gas, and the cold cloud component
from which the stars form can cool (dissipate energy) efficiently, it
drives the instability and the galactic disk fragments and forms a
number of massive clumps of stars and gas. The clumps spiral to the
center of the disk in a few dynamical times and merge there to form a
central bulge component in a strong starburst.  Noguchi (1998) had
earlier described a similar evolution in a collisional particle
simulation but, with their model for a star forming 2-phase
interstellar medium, Immeli et al.\ could both verify it in the
presence of stellar feedback effects and keep track of stellar
metallicities and ages during the evolution. This enabled them to
compute luminosities and colours for comparison with photometric
observations.  Based on these models, Immeli et al.\ (2004b) argued
that several peculiar morphological structures seen in the Hubble Deep
Field can be well explained by such a fragmented disk model, including
chain and clump cluster galaxies, and they provided predictions for
the clump masses, rotation and metallicity properties of these objects
as well as for the bulge components that form through this process.

\subsection{Bulge Properties in a Clump Merger Model}

Figure 1 shows a snapshot of the evolution in the clumpy disk phase as
well as an edge-on K-band image, the luminosity profile, the
metallicity distribution, and the [$\alpha$/iron] distribution of
bulge stars after completion of the evolution. The figure illustrates
the formation of an $\alpha$-rich bulge component in the early
evolution of this disk. For the process to work in this way the galaxy
must still be mostly gaseous at the beginning of the evolution; the
cold gas must dissipate fairly smoothly into the disk plane, before
the main star formation starts; and the clumping and star formation
must start more or less simultaneously throughout a good part of the
disk, i.e., the surface density must not be too inhomogeneous.  The
abundance of chain galaxies and clump cluster galaxies with similar
morphologies in the tadpole and ACS deep fields at redshifts around
$z\sim 2$ (Elmegreen et al.\ 2004a,b, Elmegreen \& Elmegreen 2005)
suggests that many late type and yet bulge-less galaxies may go
through such an evolutionary phase. If so, the first parts of the
Galactic bulge could well have been formed in a similar way.

\begin{figure*}[h]
   \includegraphics*[width=11.5cm,viewport=0 72 576 684, angle=-90]{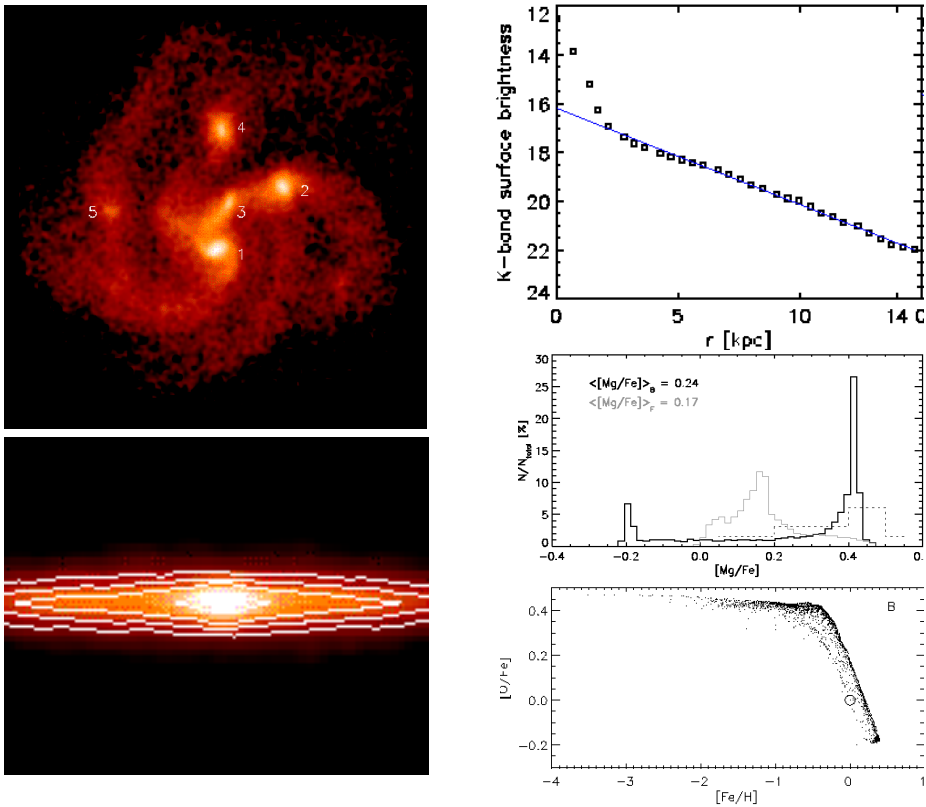}
      \caption{Left: smoothed stellar surface density of the clumpy
disk model of Immeli et al.\ (2004b) at 1.2 Gyr, with the major clumps
numbered (top), and edge-on view in the K band of the inner $11 \kpc$
of model B of Immeli et al.\ (2004a) at 3.8 Gyr, after the evolution
(bottom). Right, from Immeli et al.\ (2004a): K-band surface
brightness profile (top), distribution of [Mg/Fe] (middle) and of
[O/Fe] versus [Fe/H] for bulge stars, in model B after 4 Gyr. The
middle panel also shows for comparison the metallicity distribution of
a secularly evolving model (grey line), and the [Mg/Fe] histogram
(dotted) of 11 Galactic bulge stars measured by McWilliam \& Rich
(1994).}
       \label{figure_mafig}
   \end{figure*}



\end{document}